\documentclass{aa}
\usepackage{epsfig}
\usepackage{rotating}
\usepackage{graphicx}

\newcommand{\chan}{{\sl Chandra}}

\newcommand{\vltn}{{\em Very Large Telescope}}
\newcommand{\vlt}{{\em VLT}}
\newcommand{\hstn}{{\em  Hubble  Space Telescope}}
\newcommand{\hst}{{\em HST}}
\newcommand{\jwstn}{{\em  James Webb  Space Telescope}}
\newcommand{\jwst}{{\em JWST}}

\newcommand{\naco}{{\em NACO}}
\newcommand{\nacon}{{\em NAOS-CONICA}}

\def \psr{PSR\, B0540$-$69}

\begin{document}

\title{The near-infrared detection of PSR\, B0540$-$69 and its nebula\thanks{Based on observations collected at ESO, Paranal, under Programme 386.D-0403(A)}}

\author{R. P. Mignani\inst{1,2}
\and 
A. De Luca\inst{3,4,5}
\and
{W. Hummel}\inst{6}
\and
A. Zajczyk\inst{7,8}
\and
B. Rudak\inst{7,9}
\and
G.Kanbach\inst{10}
\and
A. S\l{}owikowska\inst{2}
}

   \institute{Mullard Space Science Laboratory, University College London, Holmbury St. Mary, Dorking, Surrey, RH5 6NT, UK
    \and
    Institute of Astronomy, University of Zielona G\'ora, Lubuska 2, 65-265, Zielona G\'ora, Poland
    \and
   INAF - Istituto di Astrofisica Spaziale e Fisica Cosmica Milano, via E. Bassini 15, 20133, Milano, Italy
 \and
 IUSS - Istituto Universitario di Studi Superiori, viale Lungo Ticino Sforza, 56, 27100, Pavia, Italy
     \and
   INFN - Istituto Nazionale di Fisica Nucleare, sezione di Pavia, via A. Bassi 6, 27100, Pavia, Italy
   \and 
   European Southern Observatory, Karl Schwarzschild-Str. 2, D-85748, Garching, Germany
   \and
   Nicolaus Copernicus Astronomical Center, ul. Rabia\'nska 8, 87100, Toru\'n, Poland
   \and
LUPM, Universit\'e Montpellier 2, CNRS/IN2P3, place E. Bataillon, 34095 Montpellier, France
\and
KAA UMK, Gagarina 11, 87-100 Toru\'n, Poland
\and
Max-Planck Institut f\"ur Extraterrestrische Physik, Giessenbachstrasse 1, 85741 Garching bei MŸnchen, Germany
    }

\titlerunning{\vlt\ observations of \psr}

\authorrunning{Mignani et al.}
\offprints{R. P. Mignani; rm2@mssl.ucl.ac.uk}

\date{Received ...; accepted ...}

\abstract{The $\sim 1700$  year old  PSR\, B0540$-$69 in the Large Magellanic Clouds (LMC)  is  considered the twin of the  Crab pulsar  because of its similar spin parameters, magnetic field, and energetics. PSR\, B0540$-$69   ($V\sim 22.5$) is also one of the very few pulsars for which both optical pulsations and polarised emission have been measured.  Its optical spectrum is fit by a power-law,  ascribed to synchrotron radiation, like for the young Crab and Vela pulsars. At variance with them, however, a  double break is required to join the X-ray and optical power-law spectra, with the first one possibly occurring in the near ultraviolet (nUV).} 
{Near-infrared (nIR) observations, never performed for \psr, are crucial to determine whether the optical power-law spectrum extends to longer wavelengths or a new break occurs, like it happens for both the Crab and Vela pulsars in the mIR, hinting at  an even more complex particle energy and density distribution  in the pulsar magnetosphere. }   
{We observed \psr\  in the J, H, and K$_{\mathrm{S}}$ bands with  the \vltn\ (\vlt)  to detect it, for the first time,  in the nIR and characterise its  optical--to--nIR spectrum. To disentangle the pulsar emission from that of its pulsar wind nebula (PWN), we obtained high-spatial resolution adaptive optics images with the \nacon\  instrument (\naco).}
{We could clearly identify \psr\ in our J, H, and K$_{\mathrm{S}}$-band images and measure its flux  (J=20.14, H=19.33, K$_{\mathrm{S}}$=18.55, with an overall error of $\pm$ 0.1 magnitudes  in each band). The joint fit to the available optical and nIR photometry with a power-law spectrum $F_{\nu} \propto \nu^{-\alpha}$ gives a spectral index $\alpha=0.70\pm 0.04$, slightly more precise than measured in the optical only. This suggests that there is no spectral break between the optical and the nIR.
We also detected, for the first time, the \psr\ PWN in the nIR. The comparison between our \naco\ images and \hstn\ (\hst) optical ones does not reveal any apparent difference in the PWN morphology as a function of wavelength. The PWN optical--to--nIR spectrum is also fit by a single power-law, with spectral index $\alpha=0.56\pm 0.03$, slightly flatter than the pulsar's. 
}
{Using \naco\ at the \vlt, we obtained the first detection of \psr\ and its PWN in the nIR. Due to the small angular scale of the PWN ($\sim 4\arcsec$) only the spatial resolution of the \jwstn\ (\jwst) will make it possible to extend the study of the pulsar and PWN spectrum towards the mid-IR. }

 \keywords{Optical: stars; neutron stars: individual:  PSR\, B0540-69}
 
   \maketitle

\section{Introduction}

Historically discovered in radio, pulsars  emit across the whole electromagnetic spectrum, from the radio, to the optical (Mignani 2011a), to the X-rays (Becker 2009),  and to high-energy $\gamma$-rays (Abdo et al.\ 2010a).  The pulsar optical emission  is due to the combination of two different processes: synchrotron radiation from relativistic electrons in the neutron star magnetosphere and thermal radiation from the cooling neutron star surface.  Their signature is recognised  by a power-law (PL; $F_{\nu} \propto \nu^{-\alpha}$)  spectrum with $\alpha\ga0$ and by a Rayleigh-Jeans (R-J) continuum, respectively. The relative weight of these two spectral components depends on the pulsar age. Young ($\la10$ kyr old) pulsars feature dominating magnetospheric emission, extending to both the  near Infrared (nIR) and the near 
ultraviolet (nUV), while the middle-aged ones ($\ga 100$ kyr  old)  feature both magnetospheric and thermal emission, with the former dominating in the nIR and the latter in the nUV.  

Very little is know about the pulsar IR emission. Only 5 out of the 12  pulsars identified in the optical, have been also detected in the nIR. These are the young pulsars Crab (e.g., Sandberg \& Sollerman 2009; Tziamtzis et al.\ 2009), Vela  (Shibanov et al.\ 2003),  and PSR\, B1509$-$58 (Kaplan \& Moon 2006),  and the middle-aged pulsars PSR\, B0656+14 and Geminga (Koptsevich et al.\ 2001).   Moreover,  the Crab and Vela pulsars have been also clearly detected in the mIR by {\em Spitzer} (Temim et al.\ 2009; Danilenko et al.\ 2011), with a possible marginal indication of detection also for Geminga (Danilenko et al.\ 2011).  {\em Spitzer} also observed  PSR\, J1124$-$5916 (Zyuzin et al.\ 2009), PSR\, J0205+6449 (Slane et al.\ 2008), and PSR\, J1833$-$1034 (Zajczyk et al.\ 2011) but it could only detect the surrounding pulsar-wind nebulae (PWNe).  
{\em Spitzer} also detected the Crab PWN, which is the only one detected both in the nIR (Sandberg \& Sollerman 2009) and in the mIR (Temim et al.\ 2009), together with the  PSR\, J1833$-$1034 PWN (Zajczyk et al.\ 2011).
A  nIR source has been recently associated with the 38.5 ms X-ray pulsar IGR\, J18490$-$0000 (Gotthelf et al.\ 2011; Ratti et al.\ 2010; Curran et al.\ 2011). However,   the identification with the pulsar has never been confirmed so far.
Detecting more pulsars in the IR is crucial to search for continuity or spectral breaks at wavelengths longer than  the optical band and, in turn, to constrain the properties of particles in the pulsar magnetosphere on a broader spectral range. Obviously, young and energetic pulsars are the best targets.

Interestingly enough, the  X-ray/radio pulsar PSR\, B0540$-$69 (Seward et al.\ 1984; Manchester et al.\ 1993) in the Large Magellanic Clouds (LMC) has not been observed yet in the IR.  With a spin-down age of $\sim 1700$  yr, it is one of the youngest pulsars so far and is similar to the Crab in its spin period ($P$=50 ms), dipolar magnetic field ($B \sim 5 \times 10^{12}$ G), and rotational energy loss rate ($\dot {E}  \sim 1.5 \times 10^{38}$ erg~s$^{-1}$). Moreover, it is the second brightest pulsar ($V\sim 22.5$) identified in the optical (Caraveo et al.\ 1992) and one of the very few for which both pulsations (Gradari et al.\ 2011) and polarisation (Mignani et al.\ 2010) have been measured in this band.  High-spatial resolution images taken with the \hstn\ (\hst) clearly resolved \psr\ from its  compact ($\sim 4\arcsec$ diameter) PWN (Mignani et al.\ 2010), allowing the most accurate measurement of the pulsar optical spectrum.  The optical PL  slope ($\alpha =0.70\pm0.07$) is similar to the X-ray one ($\alpha_{\rm X} = 0.92 \pm 0.11$; Kaaret et al.\ 2001),  but the optical fluxes fall orders of magnitudes below the X-ray PL extrapolation, as observed in most pulsars (Mignani et al.\ 2010). Interestingly enough,  while for  both the Crab and Vela pulsars a single break is sufficient to join the  optical and X-ray spectra, two breaks are required for \psr.  This unique behaviour hints at a complex particle energy and density  distribution in the pulsar magnetosphere.   \psr\ has never been observed so far in the nIR, while in the mIR the coarse spatial resolution achievable with {\em Spitzer} made it impossible to resolve the pulsar from its bright PWN (Williams et al.\ 2008). Thus, we do not know whether a single PL smoothly joins the nIR and optical spectra or  the PL slope changes once again, like it happens for the Crab and Vela pulsars in the mIR (Tziamtzis et al.\ 2009; Danilenko et al.\ 2011). In the latter case, an even more complex scenario would emerge for \psr, with  three breaks across its nIR--to--X-ray PL spectrum, a challenging one for pulsar magnetosphere models.  

Here, we present the results of the first nIR observations of \psr, recently performed by our group using adaptive optics at the \vltn\ (\vlt). This paper  is organised as follows: observations,  data reduction and analysis are  described in  Sect. 2, while  results are  presented and discussed in Sect. 3 and 4, respectively.  Conclusions follow.

\section{Observations and data reduction}

\subsection{Observation description}

\psr\  was observed in service mode on October 29th  and December 9th 2010 at the ESO  Paranal Observatory with \nacon\ (\naco), the  adaptive optics (AO) imager and  spectrometer mounted at the \vlt\ Yepun  telescope.  In order to provide  the best combination between  angular resolution and  sensitivity, we  used the  S27 camera with a pixel scale of  0\farcs027 ($28\arcsec \times28\arcsec$ field--of--view). As  a  reference  for  the  AO  correction  we  used  the  GSC-2  star S0101001160715 (Lasker et al.\ 2008), located $\sim$ 22\arcsec away from our target, with the visual ($4500-10000$ \AA) dichroic  element and wavefront sensor.    We observed \psr\ in the standard J  ($\lambda=1.265\mu$m; $\Delta \lambda= 0.25 \mu$m), H ($\lambda=1.66\mu$m; $\Delta \lambda= 0.33\mu$m) and K$_{\mathrm{S}}$ ($\lambda=2..8\mu$m; $\Delta \lambda= 0.35\mu$m) filters.   To allow for subtraction of  the variable IR  sky background and avoid possible ghosts from the bright reference star  ($V=13.05$) at the edge of the \naco\ field--of--view, we split each  integration in sequences of  short randomly dithered exposures with detector integration times (DIT) of 20 s and 24 jittered exposures (NDIT) along each node of the  5-point dithering pattern.  This corresponds to a net integration time of 2400 s per filter. 

For all  observations, the seeing   conditions measured by  the Paranal differential image motion  monitor (DIMM)   were   on   average   below   $\sim   0\farcs8$ and the airmass was $\sim$ 1.4, i.e. about the minimum possible due to the low declination of our target.  Sky  conditions  were photometric in both nights according to the ESO    ambient   condition monitor\footnote{http://archive.eso.org/asm/ambient-server}. Atmospheric water vapour content and wind speed were well within tolerance values. Night (twilight flat--fields) and daytime calibration frames (darks,  lamp flat--fields) were  taken daily  as part  of  the \naco\ calibration  plan. For both nights, standard  stars from  the Persson  et  al.\  (1998) fields were observed. 

\begin{figure}
\centering
\includegraphics[width=8.cm, angle=0,clip=]{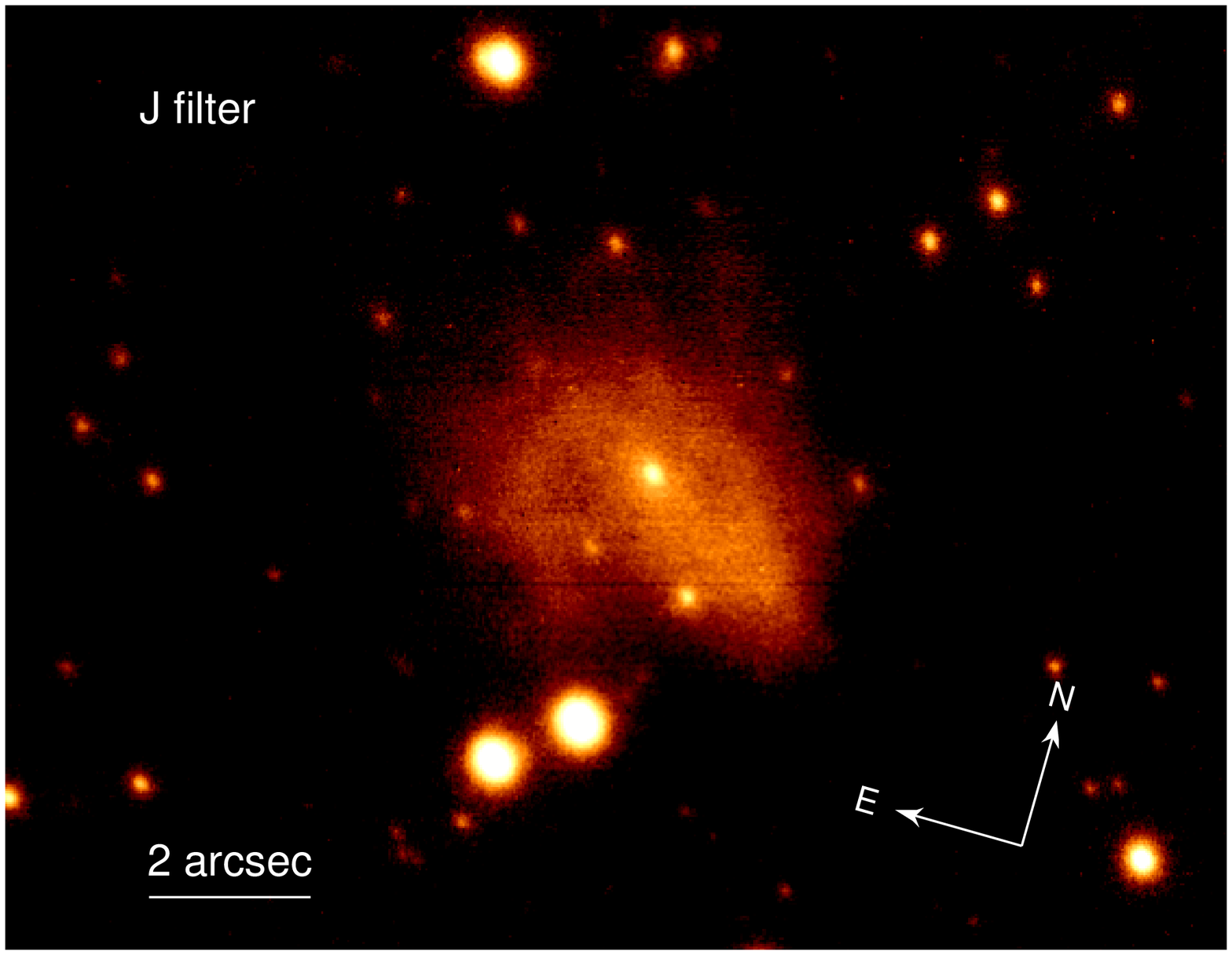}
\includegraphics[width=8.cm, angle=0,clip=]{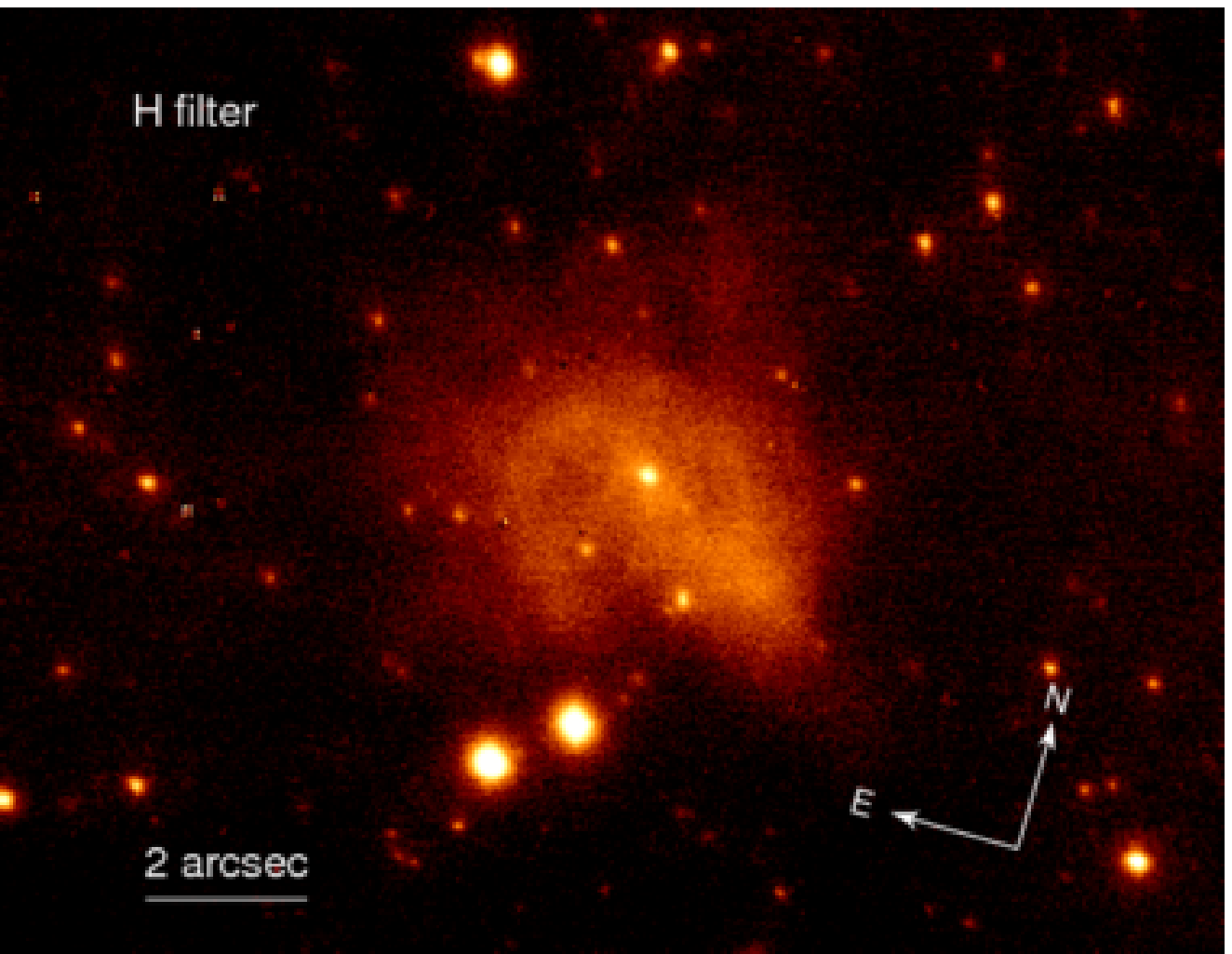}
\includegraphics[width=8.cm, angle=0,clip=]{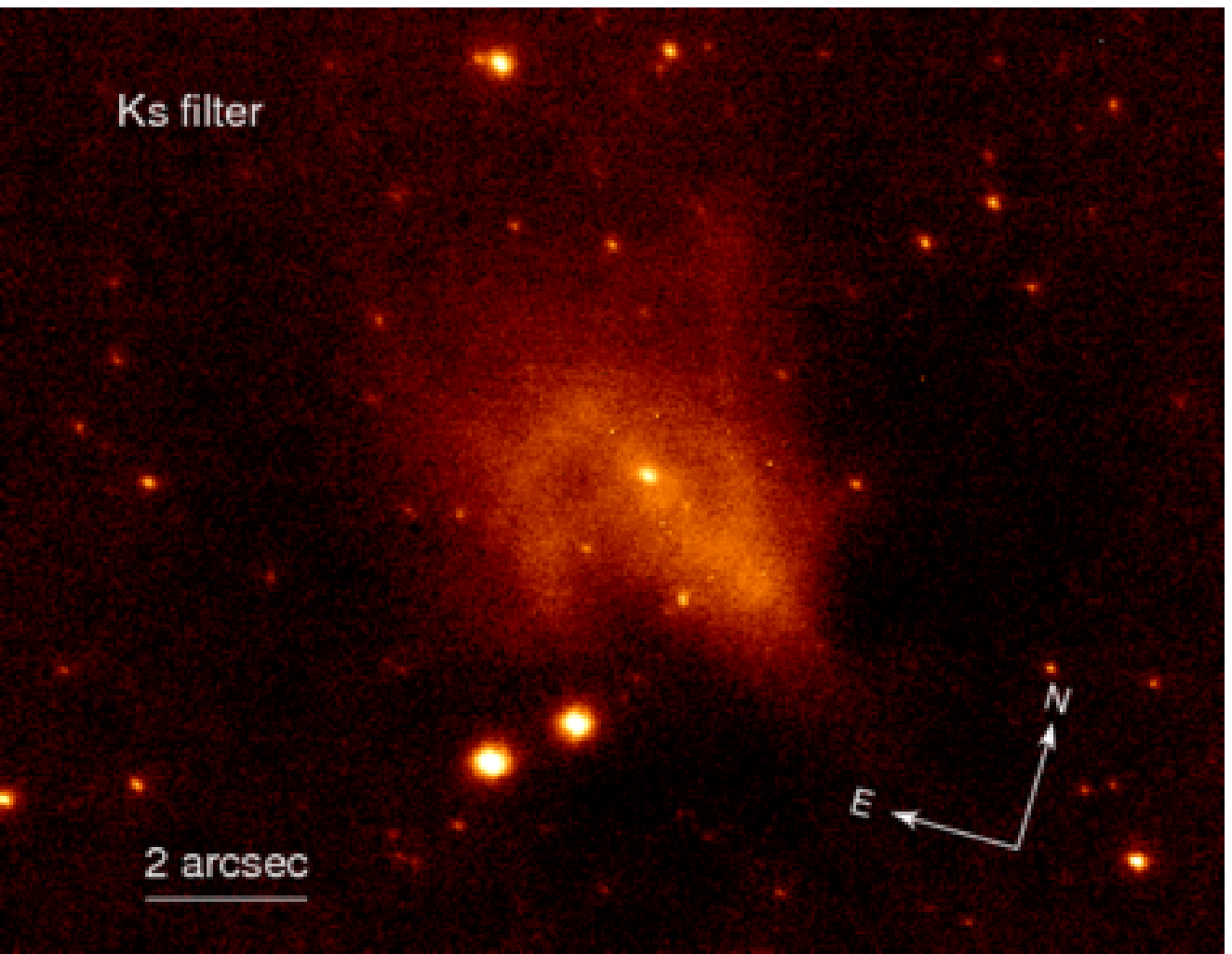}
  \caption{$15\arcsec \times 10\arcsec$ J-band image (2400 s) of the region around \psr\ and its PWN taken with \naco\ (top panel). The pulsar is the bright point source at the centre of the PWN. The H and K$_{\mathrm{S}}$-band images are also shown in the mid and bottom panels, respectively.  }
  \label{psr-fig}
\end{figure}

\subsection{Data reduction}

Calibration data  were processed   using    the   last version of the ESO   \naco\ pipeline\footnote{www.eso.org/observing/dfo/quality/NACO/pipeline}.  For each of the three filters  separately,  we dark-subtracted, flat-fielded and cleaned for hot pixels the \naco\ science images with  the produced  master dark  and flat--field frames. We also applied the correction for the odd-even column pattern affecting the \naco\ images.   When available, we also used twilight flats of similar exposure level as the sky background to minimise non-linearity effects and the fixed pattern noise. Otherwise,  we used a lamp flat.  We constructed the sky frame via a weighted median stacking of the intermediate science products. 

 Since the jitter offsets along each node of the 5-point dithering pattern are comparable to the extension of the SNR around \psr, a contamination was introduced in the computed sky frame. We corrected this contamination using the following procedure.  We decomposed the computed sky frame ${\rm S}_{\rm tot}$ into several components: the constant median value $M$, a large-scale two-dimensional gradient $G$, a component containing noise $N$, and a residual frame $F$ containing the clean sky $S_{\rm c}$ and the contamination $C$ produced by the bright SNR, with $F=S_{\rm c}+C$:

\begin{equation}
{\rm S}_{\rm tot} = M + G + N + F.
\label{datareduct}
\end{equation}

We subtracted the median $M$ and the gradient $G$from the total sky frame ${\rm S}_{\rm tot}$ and we smoothed the resulting frame using cells of  10 pixel size (0\farcs27) to separate the residual pixel-to-pixel component $N$) from the large-scale background structure to derive $F$. 
For the separation of the clean sky $S_{\rm c}$ from the contamination $C$ we replaced flux values above a certain threshold by a common fraction of their flux:

\[
  S_{\rm c} = \left\{
  \begin{array}{l l}
    R\,F &: \quad F>=T \\
    F   &: \quad F<T \\
  \end{array} \right.
\]

where the threshold $T$ and the flux scaling ratio $R$ were free parameters.  The best values for $T$ and $R$ have been found iteratively to minimise residuals of the contamination $C$ and avoid over-correction. Finally, we replaced $F$ by $S_{\rm c}$ in Eq.~\ref{datareduct},  to reconstruct the corrected sky frame.
 
We finally subtracted the cleaned sky as part of the co-adding of aligned images and  bad pixel correction procedure using the \naco\ pipeline science recipe.   The final image quality,  as measured from the median full-width half maximum (FWHM) of all star-like objects in the final product frame,  is $\sim 0\farcs16$, $\sim 0\farcs12$, and $\sim 0\farcs11$ for the J, H, and K$_{\mathrm{S}}$-band images, respectively.

As a reference for the photometric calibration we used the zero-points computed by the \naco\  pipeline, after correction for the effects of the atmospheric extinction using the most recents coefficients for the Paranal Observatory computed by  Lombardi et al.\ (2011).  Since no 2MASS stars (Skrutskie et al.\ 2006) are present in the narrow \naco/S27 field--of--view, the astrometric re-calibration of the \naco\ frames can only be computed using a grid of secondary stars selected from our \hst\ images of the pulsar field, tied to the International Celestial Reference Frame (ICRF) with a $\sim 0\farcs12$ accuracy ($1 \sigma$), as astrometric reference (Mignani et al.\ 2010). However, since we could unambiguously recognise the pulsar in the \naco\ images from relative astrometry with respect to  the \hst\ ones, we did not deem necessary to pass through this step. 

Of course, the lack of 2MASS stars in the \naco\  field--of--view also means that we have no way to counter-check our photometric calibration against an external catalogue. To rule out possible mis-calibrations, we thus checked the values of the night zero points with those taken on a 30 day time window centred on the dates of our observations and we found perfect consistency, with an rms $\sim 0.07$ magnitudes in all bands.  To rule out possible systematics in the flux calibration computed by the \naco\ pipeline, we recomputed the night zero points by matching the instrumental magnitudes of the standard stars with those in the Persson et al. (1998) catalogue. For each band, we found agreement within the rms of the \naco\ pipeline zero point distribution, after accounting for the atmospheric extinction correction. Thus, we assume $\sim 0.07$ magnitudes as the uncertainty on our absolute flux calibration, with the uncertainties associated with the  atmospheric extinction correction being negligible.

\section{Results}

The J-band image of the \naco\ field centred on the \psr\ position is shown in Fig. \ref{psr-fig} (\emph{top panel}). As seen, the pulsar is clearly resolved against its bright PWN and it is also resolved in the longer-wavelength  H and K$_{\mathrm{S}}$-band images. Thus, we detected \psr\ in the nIR for the first time and is now the sixth pulsar detected at these wavelengths.  

\subsection{Pulsar photometry}

We computed the pulsar magnitudes in the J, H, and  K$_{\mathrm{S}}$ bands  through customised aperture photometry with the IRAF task {\tt qphot}. We used an aperture of 7 pixel radius ($\sim 0\farcs19$), which is comparable with the median image FWHM and small enough to minimise the flux contamination from the surrounding, bright PWN.  Since the surface brightness distribution of the PWN is not homogeneous, and it peaks close to the pulsar, the choice of the areas used to compute the sky background is crucial for its correct characterisation and for not misestimating the pulsar flux.  We optimised the  sky background computation using an annulus located at a distance of 25 pixels (0\farcs68) from the photometry aperture, to minimise the contamination from the wings of the pulsar's PSF, and with a variable width of  20--30 pixels as a function of the filter, to account for wavelength-dependent variations in the surface brightness of the PWN. We customised the maximum size of the background annulus not to include back/foreground stars spatially coincident with the PWN.

To the pulsar fluxes measured in each filter, we then applied the aperture correction. For each filter, we computed it from the growth curve of a few bright field stars, starting from the same 7 pixel radius photometry aperture used for the pulsar. We selected those with no adjacent star within a 1\farcs5 radius and  detected not further than $\sim 5\arcsec$ from  the pulsar position to minimise aperture correction uncertainties due to the variation of the \naco\ PSF at large off-axis angles, produced by the degrade of the AO correction. Since the value of the image quality varies with the filter (see Sect. 2.1), as the result of the more efficient AO correction at longer wavelengths, it was crucial to compute the aperture correction per each filter separately. To the measured fluxes, we then applied the measured zero points and the atmospheric extinction correction. 
The pulsar magnitudes are J=20.14, H=19.33, K$_{\mathrm{S}}$=18.55, with an error of 0.1 magnitudes which accounts for the statistical errors ($\sim 0.05$), as well as for the uncertainties on our absolute flux calibration  ($\sim 0.07$) and the aperture correction estimate ($\sim 0.06$).

\subsection{The pulsar spectrum}

\begin{figure}
\centering
\includegraphics[height=8.5cm,bb=20 180 520 700, clip=]{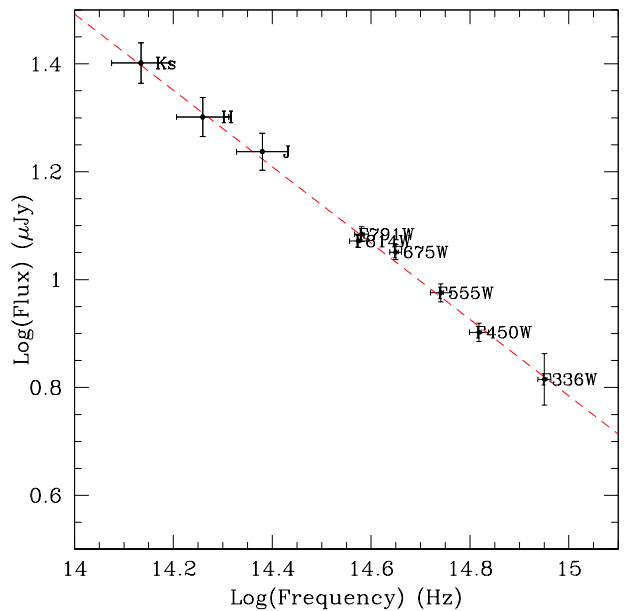}
  \caption{Extinction-corrected nIR and optical spectral flux densities of \psr.  The values in the J, H, and K$_{\mathrm{S}}$ filters have been computed from our new \naco\ observations while the values in the optical filters (labelled with numbers) are derived from the \hst\ observations of Mignani et al.\ (2010).}
  \label{psr-spec}
\end{figure}

We used our J, H, K$_{\mathrm{S}}$-band magnitudes to characterise, for the first time,  the pulsar flux in the nIR and compare its nIR spectrum with the extrapolation of the PL which best-fit the optical fluxes measured with the \hst\ (Mignani et al.\ 2010).  From the measured pulsar magnitudes we computed the spectral flux densities in each filter using as a reference the unit conversion factors from the CIT photometric system (Elias et al.\ 1982). We then corrected the flux densities for the interstellar extinction using an $E(B-V)=0.2$ (see Mignani et al.\ 2010 and references therein) and the interstellar extinction coefficients of Fitzpatrick (1999).  The results are shown in Fig.\ref{psr-spec}. As seen, the IR fluxes nicely follow the extrapolation of the optical PL spectrum ($\alpha_{\rm O} = 0.7 \pm 0.07$). A joint fit from the \hst\ 336W to the \naco\ K$_{\mathrm{S}}$-band fluxes give a spectral index $\alpha_{\rm O,nIR} = 0.70 \pm 0.04$. For comparison, a fit to the nIR data points alone would yield a spectral index $\alpha_{nIR} = 0.66 \pm 0.2$, i.e. fully compatible with that measured across the whole spectral range. Thus, we found no evidence of spectral breaks between the optical and nIR, at variance with what is expected between the optical and the X-rays.  Our derived spectral index is virtually identical to that measured by Mignani et al.\ (2010) using only the \hst\ data. However, thanks to the larger wavelength coverage, our spectral fit is slightly  more accurate, although the larger error bars on the \naco\ photometry do not allow to have a more significant improvement. 

Our data obviously do not rule out the presence of a spectral break towards longer wavelengths like that found,  for instance,  in the spectrum of the Crab and Vela pulsars (Temim et al.\ 2009; Danilenko et al.\ 2011). Observations in the mIR would be crucial to search for such break but, unfortunately, the coarse angular resolution of {\em Spitzer} does not allow one to resolve the pulsar emission against that of its PWN.  Moreover, the  pulsar+PWN {\em Spitzer} fluxes in the IRAC1--4 (3.6, 4.5, 5.8, and 8.0 $\mu m$) and MIPS1 ($24 \mu m$) channels (Williams et al.\ 2008) are about an order of magnitude above the extrapolation of the optical--to--nIR PL spectrum of the pulsar, so that they are obviously non constraining.  A spectral break at wavelength as long as $5\mu m$  in the \psr\ spectrum can only be detected with the Near Infrared Camera (NIRCam) aboard the \emph{James Webb Space Telescope} (\emph{JWST}),  which has an high enough spatial resolution required to resolve the pulsar emission from that of its PWN.

\subsection{The nebula morphology}

\begin{figure}
\centering
\includegraphics[scale=0.6]{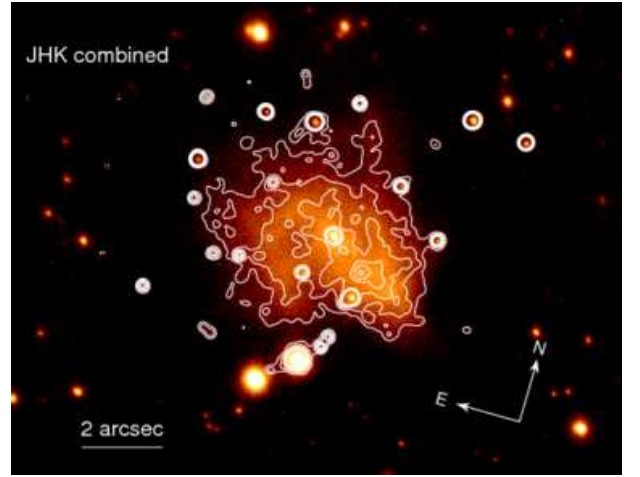}
 \caption{Co-added JHK$_{\mathrm{S}}$-band image of the \psr\ PWN with the contours of the 1995 \hst/WFPC2 F555W image (Mignani et al.\ 2010) overlaid. Note that the emission knot at $\sim2\farcs2$ SW of the pulsar, visible in the \hst/WFPC2 contours (epoch 1995.87), is not detected in our more recent (epoch 2010.94) \naco\ images.}
 \label{pwn-fig}
\end{figure}

We clearly detected the \psr\ PWN in the J, H and K$_{\mathrm{S}}$ band images (see Fig. \ref{psr-fig}) and, it is now the fifth PWN detected in either the nIR or mIR.  Similarly, as for the pulsar, the PWN is detected for the first time in the nIR.  We studied both the morphology of the PWN in the different nIR bands and its nIR spectrum.

As seen from Fig. \ref{psr-fig}, there are no obvious differences in the nebula structure seen in different nIR bands. The size of the PWN in the nIR is similar to that seen at shorter wavelengths, as seen from the comparison with the \hst/Wide Field Planetary Camera 2 (WFPC2)  images of Mignani et al.\ (2010). Thus, no variation in the nebula size is observable when going  from the optical to the nIR  (Fig. \ref{pwn-fig}).  Like in the optical (e.g., Serafimovich et al.\ 2004), we can distinguish two main structures in the PWN. The first one is a bright bar, along a position angle PA$\sim 45^{\circ}$ (measured North through East), that stretches at both sides of the pulsar and is identified as a probable torus (Gotthelf \& Wang 2000; Serafimovich et al.\ 2004).  Superimposed to the bar, south-west (SW) to the pulsar, there is  a blob of bright  emission. This feature is also visible in the optical \hst\ data (De Luca et al.\ 2007) and also in the X-ray \chan\ data (Gotthelf \& Wang 2000). However, the brightness contrast between the blob and  the north-east (NE)  side of the putative torus seems to be more pronounced in the X-rays than in the nIR or in the optical. To the north of the pulsar, a weak slightly extended emission is recognised, also seen in the optical data. Most probably, this feature is associated with a possible jet identified in the optical by Serafimovich et al.\ (2004). Like in the optical, we observe differences in the surface brightness on small angular scales ($\la 1\farcs0$), which might imply differences in the injection of relativistic particles in the PWN, possibly due to a different wind geometry. Alternatively, they might be produced when Rayleigh-Taylor instabilities occur at the interface between the PWN bubble and the surrounding cold SNR ejecta producing, e.g. filaments and/or fingers.  We could not detect the bright emission knot observed in the optical  by the \hst\ $\sim2\farcs2$ SW of the pulsar (Luca et al.\ (2007).  This can be explained either by a sharp break in the spectrum of the knot or, more likely, by its intrinsic variability, possibly associated with a change in its apparent position. Indeed, both the fading of the knot and its apparent displacement are clearly noticed from the comparison between the  multi-epoch \hst\ observations of De Luca et al.\  (2007) and Mignani et al.\  (2010), which span a time base line of $\sim 12$ years.  Possible scenarios  on both the knot origin and its fading are discussed in Lundqvist et al.\ (2011).

\subsection{The nebula spectrum}

\begin{figure}
 \centering
 \includegraphics[scale=0.6]{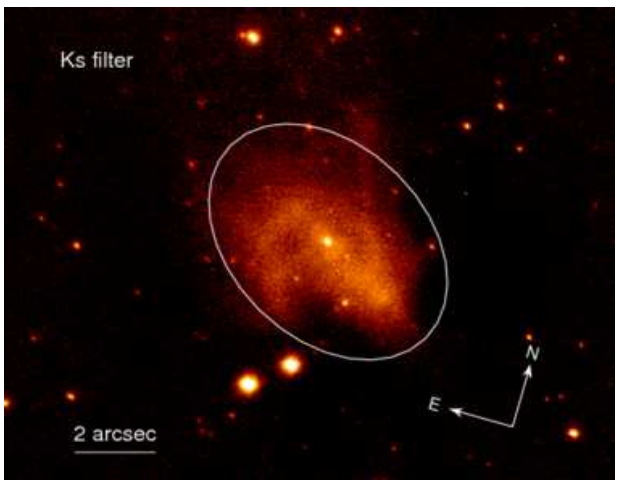}
 \caption{K$_{\mathrm{S}}$-band image of the \psr. The extraction region used for estimation	of the PWN flux in J, H, K$_{\mathrm{S}}$ bands and also in the \emph{HST} bands (Mignani et al.\ 2010) is shown with an ellipse.  }
 \label{pwn-ellipse}
\end{figure}

We used our \naco\ data to characterise the optical/nIR spectrum of the PWN.
We measured the JHK$_{\mathrm{S}}$ fluxes for the whole PWN using an elliptical extraction region of size $3\farcs4 \times 2\farcs4$ for semi-major and semi-minor axes, respectively (see Fig. \ref{pwn-ellipse}).
The contribution from the pulsar and stars overlapping the PWN extraction regions was evaluated through aperture photometry and subtracted from the total flux.  
The correction for the airmass and interstellar extinction has been applied as for the pulsar (see Sectn.\ 3.2). Using the same extraction regions as for the JHK$_{\mathrm{S}}$ bands, we also calculated the corresponding optical fluxes using the  \emph{HST/WFPC2} data of  Mignani et al.\ (2010). As before, the contribution from the pulsar and overlapping stars were subtracted from the measured values. Following the \emph{WFPC2} Data Handbook (Bagget  et al.\ 2002) and the approach described in Sect.~4.2. of Mignani et al.\ (2010) we corrected the fluxes for the charge transfer efficiency losses of the \emph{WFPC2} detector. Correction for the interstellar extinction was applied as usual. 
The final results are summarised in Table\ 1.
We then fitted the intrinsic fluxes of the whole PWN with a PL  $F_{\nu} \propto \nu^{-\alpha}$. Like for the pulsar, we computed the fitting with a least-squares fitting method that  takes into account the uncertainties of the measured fluxes as their weights. This yields a PWN spectral index of $\alpha^{\rm PWN}_{O,nIR}= 0.56 \pm 0.03$ for the data in the nIR/optical range.

\begin{table}
\begin{center}
\caption{Extinction-corrected fluxes for the \psr\ PWN. Fluxes have been measured for the \vlt\ J, H and K$_{\mathrm{S}}$ bands (first three rows) and for the \emph{HST}$^{*}$ data presented in Mignani et al.\ (2010).}
\begin{tabular}{llc} \hline\hline
Telescope/ & Band	  & Flux \\
Instrument	          &               &[$10^{-27}$ erg s$^{-1}$ cm$^{-2}$ Hz$^{-1}$]\\ \hline	  
	         \vlt/\naco\ & J & $5.41 \pm 0.22$ \\
	                   & H & $6.75 \pm 0.27$ \\
	                   & K$_{\mathrm{S}}$ & $8.01 \pm 0.35$ \\ \hline
	 \hst/{\em WFPC2} & F336W & $2.11	\pm 0.22$ \\
	  &F450W & $3.61	\pm 0.22$ \\
	  &F555W & $3.46	\pm 0.17$ \\
	  &F675W & $4.72	\pm 0.18$ \\
	  &F814W & $4.81	\pm 0.14$ \\
	  &F791W & $4.26 \pm 0.13$  \\ \hline
	   \end{tabular}
	   \label{flux-tbl}
 \end{center}
  $^{*}$All the \emph{HST} data were acquired in 2007, except for the data in the F791W filter that were taken in 1999.
\end{table}

\begin{figure}[h]
\centering
\includegraphics[bb=0 20 541 445, scale=0.6, angle=0,clip=]{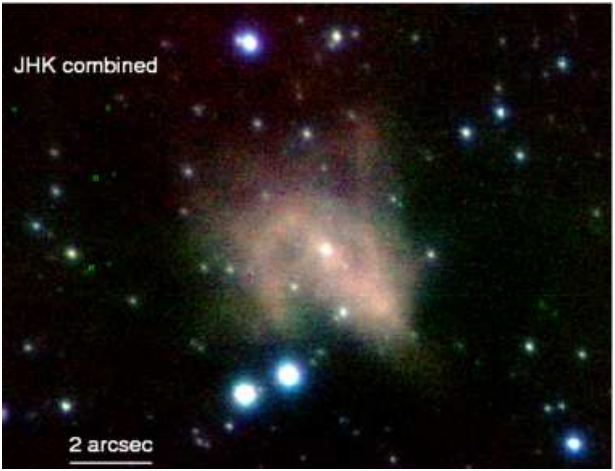}
 \caption{Co-added JHK$_{\mathrm{S}}$-band image of the \psr\ nebula. Images in different bands have been colour-coded (J=blue, H=green, K$_{\mathrm{S}}$=red).}
 \label{pwn-rgb}
\end{figure}

We also qualitatively searched for evidence of possible variations in the nIR spectrum of the PWN as a function of position on angular scales smaller than $\sim 1\arcsec$.   We note that  a systematic work done in the optical using \hst\ data shows that the PL spectrum of the PWN does not change significantly when different positions  are sampled both around the bar and around the putative jet and counter-jets (see, e.g. Serafimovich et al.\ 2004;  Lundqvist et al.\ 2011).  Fig. \ref{pwn-rgb} shows the composite JHK$_{\mathrm{S}}$-band image of the \psr\ PWN, where the emission contributions in different bands are colour-coded. As seen, we cannot find evidence of regions more prominent in a band than in another. Thus, we conclude that there are no significant small-scale variations in the nIR spectrum of the PWN, like in the optical. We note, however, that the intrinsic variability of the PWN on angular scales of $\la 1\arcsec$ (e.g., De Luca et al.\ 2007) can affect the local multi-band spectral analysis. For this reason, the comparison of multi-band observations taken years apart must be taken with due caution. Indeed, while our \naco\ data have been taken within few weeks from each other (Sectn.\ 2.1), the available \hst\ data are about 3.5 to 15 years older (Mignani et al.\ 2010) and, in principle, are not directly comparable with the former.  In the X-rays, a spectral analysis of different regions of the PWN has been performed using \chan\  data (Petre et al.\ 2007).  A quite complete discussion of the comparison between the spatially-resolved optical and X-ray spectra of the PWN is given both in Serafimovich et al.\ (2004) and Lundqvist et al.\ (2011), and we will not repeat it here.

\section{Discussion}

\subsection{The pulsar infrared luminosity}

\begin{figure}
\centering
\includegraphics[height=8.5cm,bb=20 180 520 700, clip=]{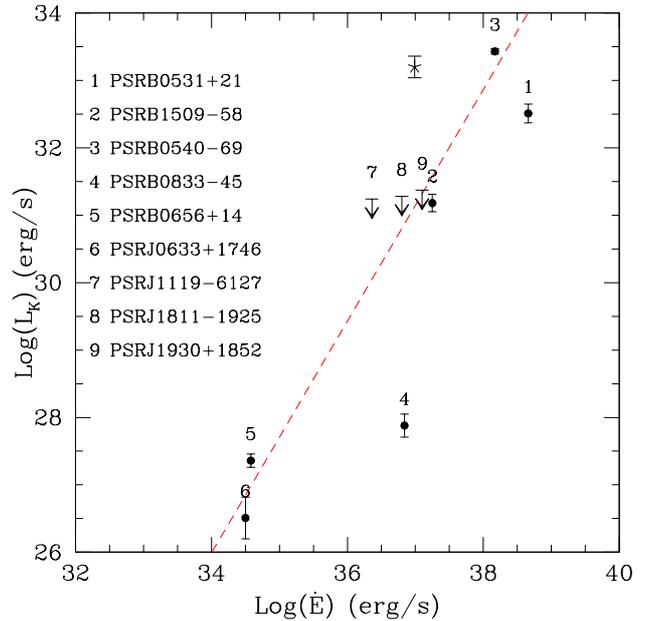}
  \caption{Measured K-band luminosities of identified rotation-powered pulsars (filled circles) and available upper limits plotted as a function of the rotational energy loss rate $\dot{E}$ (updated from Mignani et al.\ 2007b). Different pulsars are labelled accordingly (see legenda). The plotted errors on the K-band luminosity account for the uncertainties on the flux measurements, the extinction correction, and the nominal pulsar distance. For \psr, the error is comparable with the size of the symbol.   The star indicates the candidate nIR counterpart to the X-ray pulsar IGR\, J18490$-$0000 (Ratti et al.\ 2010; Curran et al.\ 2011). }
 \label{psr-lum}
\end{figure}

We computed the \psr\ luminosity by using as a reference the most recent value of the LMC distance ($48.97\pm 0.9$ kpc), obtained from re-calibrating the Cepheids period-luminosity relation (Storm et al.\ 2011). At the assumed distance, the measured extinction-corrected K-band flux of \psr\ corresponds to a luminosity $L_{K} \sim 2.8 \times 10^{33}$ erg s$^{-1}$.  We plotted in Fig.~\ref{psr-lum} the K-band luminosity of all rotation-powered pulsars as a functional of the rotational energy loss rate $\dot{E}$.  We have corrected the observed K-band fluxes using the interstellar reddening $A_V$ measured along the line of sight (see Mignani et al.\ 2007a and references therein) and the interstellar extinction coefficients of Fitzpatrick (1999). We assumed, the best available measurements of the pulsar distances, either from optical/radio parallax or from the DM. We also plotted  the measured upper limits for other pulsars observed in the nIR, i.e. PSR\, J1119$-$6127 (Mignani et al. \ 2007a), PSR\, J1811$-$1925 and  PSR\, J1930+1852 (Kaplan \& Moon 2006).  As seen, the K-band luminosity correlates with the $\dot{E}$, with $L_{K} \propto \dot{E}^{1.72\pm0.03}$. The available luminosity upper limits are also consistent with such a trend. The slope of the $L_{K}$--$\dot{E}$ relation is somewhat steeper than previously measured by Mignani et al.\  (2007a), i.e. $L_{K} \propto \dot{E}^{1.3\pm0.04}$, when \psr\ was yet to be detected in the nIR.  The former is equal to the corresponding slope measured in the optical band for the same set of pulsars, where $L_{V} \propto \dot{E}^{1.70\pm0.03}$.   
Notwithstanding all the associated uncertainties and the very limited statistics, the observed correlation between the nIR luminosity and $\dot{E}$  clearly clearly suggests that the pulsar nIR emission is  rotation-powered and rules that it is entirely produced by a debris disk.  This emission may then be of either purely magnetospheric origin or partly coming from a pulsar's wind termination shock region.

As seen from  Fig.~\ref{psr-lum}, \psr\ is the most luminous pulsar in the nIR. For its  rotational energy loss rate  ($\dot {E}  \sim 1.5 \times 10^{38}$ erg~s$^{-1}$), the derived nIR luminosity corresponds to an emission efficiency $\eta_{nIR} \equiv  L_{K}/\dot{E} \approx 10^{-5}$.  This is higher than the emission efficiencies computed for all the other pulsars detected in the nIR, for which  $\eta \sim 10^{-9}$--$10^{-6}$ (see also Mignani et al.\ 2007a). In particular, \psr\ turns out to be about an order of magnitude more efficient that its "twin", the Crab pulsar,  whose location in Fig.~\ref{psr-lum} is slightly below the luminosity-$\dot{E}$ trend.   
Although the measured flux of \psr\ is sensitive to the choice of the background area (see Sectn. 3.1), it is clear that the difference in the inferred nIR  emission efficiency  is too large to be ascribed to an over-estimate of the pulsar flux due to an inaccurate subtraction of the nebula background.  We note that a larger emission efficiency of \psr\ with respect to the Crab is also observed in the optical (e.g., Zharikov et al.\ 2006).
Thus, it has to be related to a difference in the intrinsic pulsar emission properties, both in the nIR and in the optical. Non-thermal activity of young pulsars like the Crab and \psr\ is  due to spatially extended outer gaps in their magnetospheres. Unfortunately, there exists no well-established self-consistent model of outer gaps which may be applicable even to pulsars of similar $P$ and $\dot P$. It is then difficult to explain, for instance, the difference between \psr\ and the Crab pulsar in $\gamma$-rays. The same remains true for the optical to nIR emission. As seen from Fig.~\ref{psr-lum}, the Vela pulsar, is clearly under-luminous  with respect to the luminosity--$\dot{E}$ trend,  which implies an intrinsically lower emission efficiency $\eta_{nIR} \approx 10^{-9}$.  A similarly low emission efficiency is also observed in the optical (Zharikov et al.\ 2006). Like for the Crab, this must be related to a difference in the intrinsic pulsar emission properties in these spectral ranges.

We also compared the inferred nIR luminosity of the candidate counterpart to the X-ray pulsar IGR\, J18490$-$0000 (Ratti et al.\ 2010; Curran et al.\ 2011) with those of the pulsars identified in the nIR. As a reference to compute the IGR\, J18490$-$0000 nIR luminosity, we used its observed magnitude K$_{\mathrm{S}}=17.2\pm0.4$ and interstellar reddening $E(B-V)=6.62$ (Curran et al.\ 2011) and we assumed a distance of 7 kpc, as in Gotthelf et al.\  (2011).  The location of the IGR\, J18490$-$0000 candidate nIR counterpart in Fig.~\ref{psr-lum} is well above the luminosity-$\dot{E}$ trend, while its inferred nIR luminosity is comparable to that of \psr\ but for a factor of 10 lower $\dot{E}$ (Gotthelf et al.\ 2011). This would imply a nIR emission efficiency for  IGR\, J18490$-$0000 which is a factor of 10 larger than \psr. Such a value is surprisingly high, even accounting for the large uncertainties on the distance and interstellar extinction towards IGR\, J18490$-$0000 (Gotthelf et al.\ 2011). The inferred efficiency would be even higher if one assumes the observed magnitude (K$_{\mathrm{S}}=16.4\pm0.1$) of the candidate counterpart originally reported in Ratti et al.\ (2010). This might argue against the proposed identification, suggesting that the candidate nIR counterpart is an unrelated field star, which is plausible according to the relatively high ($\sim 6\%$) chance coincidence probability\footnote{A much lower probability ($\sim 1.6 \times 10^{-5}$), however,
   is quoted by Ratti et al.\ (2010), although they do not explain how
   they derived this value.}  in the field (Curran et al.\ 2011). New nIR observations are required to obtain a more precise estimate of the distance and the extinction towards IGR\, J18490$-$0000 and determine the spectrum of its candidate counterpart.

\subsection{The pulsar multi-wavelength spectrum }

\begin{figure}
\centering
\includegraphics[height=9.0cm, angle=270,clip=]{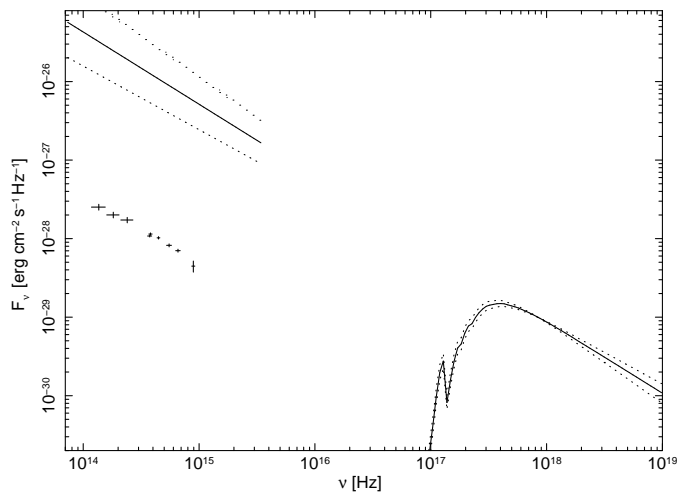}
  \caption{Optical-to-nIR spectral energy distribution of \psr\  (points) compared with the PL model (Kaaret et al.\ 2001) best-fitting the \chan\ X-ray spectrum (solid line) and its extrapolation in the optical domain. Dashed lines correspond to a $1 \sigma$ uncertainty on the model parameters. }
  \label{psr-mw}
\end{figure}

The  multi-wavelength spectral energy distribution (SED) of \psr\ is shown in Fig.\ref{psr-mw}, where the optical--to--nIR PL is compared with that best-fitting the \chan\ spectrum (Kaaret et al.\ 2001) and its extrapolation in the optical domain.  Due to the agreement between the optical and nIR spectra, the overall SED picture does not change with respect to that described in Mignani et al.\ (2010), with the spectral index of the optical--to--nIR PL ($\alpha_{\rm O,nIR}=0.70\pm 0.04$) being quite similar to the
X-ray one obtained from phase-integrated spectroscopy ($\alpha_{\rm X}=0.92 \pm 0.11$),  and  the  optical--to--nIR  spectrum undershooting by a  factor  of  100  the  low-energy extrapolation of the 0.1-10 keV X-ray spectrum.  The peculiarity of the \psr\ SED, with the presence of two spectral breaks,  has been already noted by Serafimovich  et al.\ (2004) and Mignani et al.\ (2010), and we refer to the discussion therein.  Here, we compare the optical--to--nIR spectrum of \psr\ with those of other pulsars detected in both wavelength ranges.

Although the correlation with the $\dot{E}$ (Sectn.\ 4.1) proves that the pulsar nIR emission is rotation-powered, all pulsars feature markedly different behaviours in the nIR/mIR with respect to the optical/nUV, which makes it impossible to fit the overall spectrum with a single-component  model.
For instance, the Crab pulsar shows a clear spectral turnover red-ward of the nUV, where the PL spectral index $\alpha_{\rm nUV}$ gradually turns from positive into negative, smoothly joining the optical/nIR spectrum (see, e.g. Fig.\ 4 of Mignani et al.\ 2010), but features a new change in the mIR, where it becomes approximately zero  (Temim et al.\ 2009). For the Vela pulsar, instead, a single PL  fits the flat nUV--to--nIR spectrum (Kargaltsev \& Pavlov 2007) but the PL  spectral index  turns into positive at wavelengths longer than 2.2 $\mu$m, producing a spectacular rise of the PL spectrum in the mIR (Danilenko et al.\ 2011). For PSR\, B1509$-$58, the lack of multi-band detections at optical wavelengths, with only an R-band flux measurement obtained from polarimetry observations and, probably, affected by the presence of a brighter, nearby, star (Wagner \& Seyfert 2000), does not allow to draw any definite  conclusion. Nonetheless, the existing flux measurements are consistent with a single PL with positive spectral index fitting the optical/nIR spectrum. On the other hand, for PSR\, B0656+14 the composite optical/nUV spectrum sharply turns into a PL  with positive spectral index in the nIR (Koptsevich et al.\ 2001), while for Geminga it shows a more gradual but similar turnover (Koptsevich et al.\ 2001) which seems to smoothly extend to the mIR (Danilenko et al.\ 2011).   However, the cases of PSR\ B0656+14 and Geminga are different from those of the younger pulsar, since the observed spectral turnover between the optical/nUV and the nIR purely marks the transition between a Rayleigh-Jeans and a synchrotron-dominated continuum, and not between different slopes of the synchrotron PL spectra.   

Thus, the spectra of the Crab and Vela pulsars suggest that PL spectral  breaks between the optical/nIR and the mIR can be a characteristic of the pulsar magnetospheric emission. This might be true, in principle, also for \psr\ and PSR\, B1509$-$58 for which, however, no mIR observations are available yet. Similar breaks are also observed between the optical/UV and the X-ray bands (Mignani et al.\ 2007b) and are likely related to the different energy and density distribution of particles in different regions of the pulsar magnetosphere, where the optical and X-ray radiation is produced.  The presence of spectral breaks also between the optical/nIR and the mIR suggests rather different emission regions for the observed radiation, while, on the other hand, the continuity between the optical and nIR suggests that it is produced in very close regions in the neutron star magnetosphere. The comparison of the pulsar light curves in the optical and in the nIR/mIR would be crucial to verify this interpretation, since differences in the light curve morphology and in the peak alignment would automatically reflect different locations of the emission regions.  However,  only very few pulsars have been observed to pulsate in the optical (see Mignani 2010, 2011b for recent reviews),  including the Crab and Vela pulsars and \psr. Of them, only the Crab (Eikenberry et al.\ 1997) has been observed to pulsate in the nIR,
although not yet with newest-generation instruments. For the Crab, the pulsar light curve evolves from the optical to the nIR, with evidence of an increase of the peak-to-peak phase separation and width as a function of wavelength, suggesting a different geometry of the emission regions.  No pulsar has been searched yet for pulsations in the mIR.  

The level and spectral shape of intrinsic magnetospheric emission in outer gap models depends strongly on many factors which may vary from pulsar to pulsar. In particular, the details of the gap structure and its boundary conditions may be different, leading in turn to quite different properties of the non-thermal pulsar radiation.
This is equally true in the high-energy domain (from X-rays to $\gamma$-rays) and in the low-energy domain (from UV to optical, to nIR and to longer wavelengths). The most sophisticated 3D model to date has been worked out for the Crab pulsar only (Hirotani 2008). However, even this model has too many free parameters and, therefore, is subject to  continuous modifications driven by new discoveries (Section 8.2 in Aleksic et al.\ 2011). Any high-quality photometric and/or spectral data in the low-energy range may shed light on  the location and spatial extent of the regions of origin of emission which 
in outer-gap models comes from high-order generation cascades inside and outside the gaps.

\subsection{The PWN multi-wavelength spectrum}

\begin{figure*}[t]
 \centering
  \begin{tabular}{cc}
  \includegraphics[bb=70 360 558 720,scale=0.7,clip=]{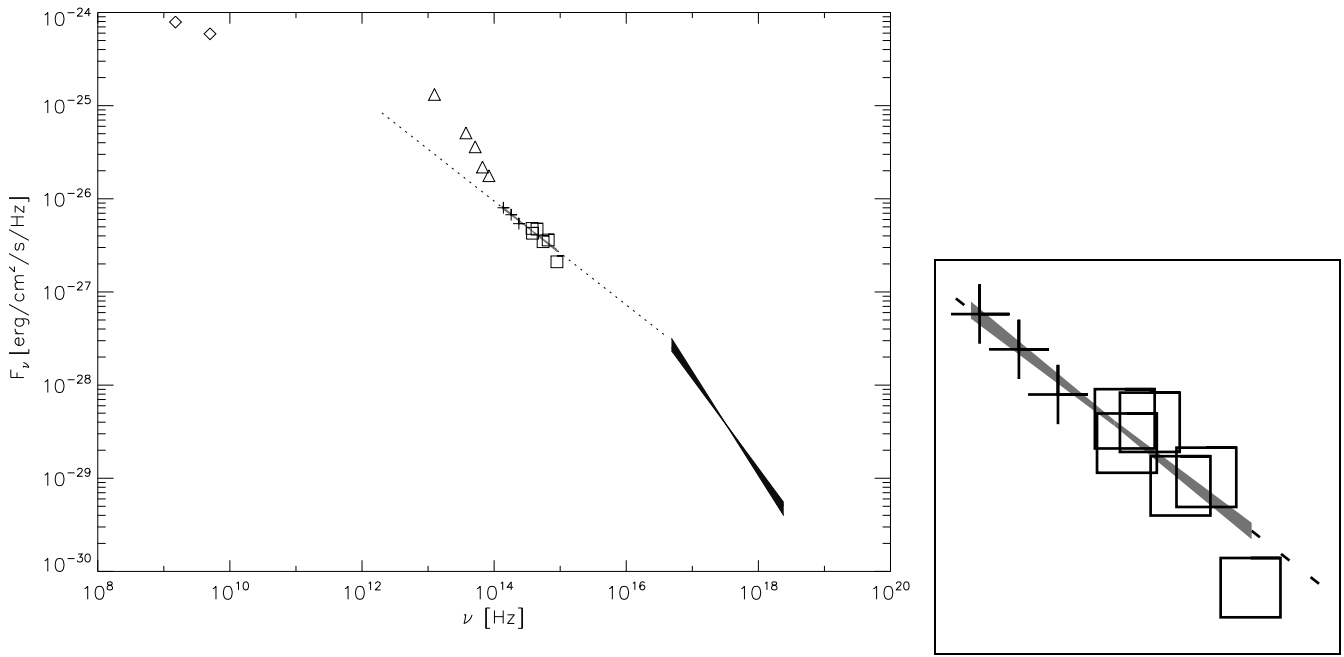}
 \framebox{ \includegraphics[bb=330 560 370 600,height=5cm,clip=]{19177_f10.eps}}
  \end{tabular}
 \caption{SED of the \psr\ PWN with our \naco\ JHK$_{\mathrm{S}}$ fluxes (crosses), the \hst\ optical fluxes (rectangles) obtained from the data of Mignani et al.\ (2010), the \emph{Spitzer} IRAC and MIPS fluxes (triangles) of Williams et al.\ (2008), and  the radio fluxes (diamonds) from Manchester et al.\ (1993). The \chan\ PL fit to the X-ray spectrum of the PWN (Kaaret et al.\ 2001) is shown as a black bow-tie. The PL fit to the IR and optical data is shown as a dark grey bow-tie (see zoom on the right panel).
	The dotted line shows the extrapolation of the optical/nIR spectrum towards the X-ray and the mIR.}
 \label{pwn-spec}
\end{figure*}

Our \naco\ detection of the PWN around \psr\ makes it one of the very few PWN systems detected both in the optical and either in the nIR or mIR. The others are the Crab (Sandberg \& Sollerman 2009; Temim et al.\ 2009), PSR\, J0205+6449  (Shibanov et al.\ 2008; Slane et al.\ 2008), and PSR\, J1124$-$5916  (Zharikov et al.\ 2008; Zyuzin et al.\ 2009).  No other PWNe  are detected so far in either of these spectral ranges, with the only exception of the PSR\, J1833$-$1034 PWN, which has been detected both in the nIR and mIR with the \vlt\ and {\emph Spitzer}, respectively (Zajczyk et al.\ 2011).  In all cases, the PWNe are also detected in the X-rays, where they have similar morphologies as in the optical/IR, although, in some cases, with a somewhat larger spatial extent  (e.g., the PSR\, J0205+6449 PWN). The relation between the optical/IR and X-ray spectra of these PWN, however, is not unique.  For instance, for both the Crab and the PSR\, J0205+6449 PWNe,  the nebula spectrum is compatible with a single PL, extending from the X-rays to the optical/IR. On the other hand, for the PSR\, J1124$-$5916 PWN,  the optical/IR spectrum clearly undershoots the extrapolation of the X-ray one, implying a break in the multi-wavelength PL spectrum between the optical/IR and the X-rays bands.

Fig.~\ref{pwn-spec} shows the SED of the \psr\ PWN. Interestingly, the optical/nIR PL spectrum ($\alpha^{\rm PWN}_{\rm O,nIR}= 0.56 \pm 0.03$) is flatter than  the pulsar ($\alpha_{\rm O,nIR}=0.70\pm 0.04$), suggesting that the emission becomes harder at large distances from the pulsar. We note that the PWN spectral index in the optical/nIR  is flatter than measured in the optical by Serafimovich et al.\ (2004) based on a subset of our \hst\ data set. This is probably due to the 336W flux, which is lower than the best-fit optical/nIR PL and, then, produces a steepening of the PL in the optical.   The best-fit  \emph{Chandra} PL spectrum of the PWN in $0.6$--10~keV energy range Kaaret et al.\ (2001) is also shown in Fig.\ref{pwn-spec}, together with the mIR fluxes from the \emph{Spitzer} IRAC and MIPS observations (Williams et al.\ 2008) and the radio fluxes from Manchester et al.\  (1993). The extrapolation of the PWN X-ray spectrum, with spectral index $\alpha^{\rm PWN}_{\mathrm{X}} = 1.04$,  to the nIR-optical domain clearly overestimates the PWN optical and nIR fluxes. This suggests that a spectral break must exist somewhere between the optical and  X-ray bands. This behaviour is at contrast with what is observed in this wavelength range for \psr, where two spectral breaks are needed to match the pulsar PL spectrum between the optical and X-rays (see Sect.\ 4.2). 
When extrapolating the PWN nIR spectrum ($\alpha^{\rm PWN}_{\rm O,nIR} = 0.56\pm 0.03$) to the mIR, it is obvious that the fitted spectrum underestimates the \emph{Spitzer} fluxes. However, word of caution is necessary before drawing the conclusion that the spectrum hardens between nIR and mIR bands. As described in Williams et al.\ (2008) the IRAC and MIPS mIR fluxes were estimated using rather large circular aperture (radius of $6\arcsec$) compared to the size of the elliptical aperture used in our calculations (see Fig.\ref{pwn-ellipse}). Moreover, our  nIR and the optical fluxes were corrected for the contribution from the pulsar and the background/foreground stars overlapping the PWN, while this is clearly not the case for the \emph{Spitzer} fluxes, where both the pulsar and the stars are not resolved at the angular scale of the IRAC and MIPS data. For this reason, the true intrinsic nebular fluxes in the mIR  are, most likely, lower that those presented in Williams et al.\ (2008).
We note that the PWN spectral index has to change once more somewhere between the nIR/mIR and the radio bands from $\alpha^{\rm PWN}_{\rm nIR} = 0.56$ to $\alpha^{\rm PWN}_{\mathrm{R}} = 0.25 \pm 0.1$ (Manchester et al.\ 1993). A similar spectral turnover  towards radio frequencies is also observed for the Crab (e.g., review by Hester 2008; Serafimovich et al.\ 2004)  and for the PSR\, J0205+6449 (Slane et al.\ 2008) PWNe.

\section{Summary and conclusions}

Thanks to high-spatial resolution observations performed with \naco\ at the \vlt, we  obtained the first detection of \psr\ and its PWN in the nIR. The pulsar's broad-band photometry fluxes in the optical (Mignani et al.\ 2010) and nIR are fit by a single PL with spectral index $\alpha_{\rm O,nIR}=0.70\pm 0.04$, extending from $\sim 0.3$ to $2.8 \mu m$. This implies continuity between the optical and the nIR spectra, although we cannot rule out yet the presence of a spectral break towards the mIR, like in the case of the Crab and Vela pulsars. The K$_{\mathrm{S}}$-band luminosity of \psr\ ($L_{K} \sim 2.8 \times 10^{33}$ erg s$^{-1}$) well correlates with the rotational energy loss $\dot{E}$, like in other pulsars, confirming the non-thermal nature of the nIR emission. In particular, at the LMC distance, \psr\ is the pulsar with the largest nIR luminosity, about an order of magnitude larger than the Crab's. This  implies a correspondingly higher emission efficiency, like in the optical (e.g., Zharikov et al.\ 2006). The PWN has both a nIR morphology  and a spatial extent consistent with those observed in the optical with the \hst. We could not detect the bright emission "knot" observed in the \hst\ images SW of the pulsar (De Luca et al.\ 2007). This can be explained either by a sharp break in its spectrum or, more likely, by its intrinsic variability, as shown by  the multi-epoch \hst\ observations (De Luca et al.\ 2007; Mignani et al.\ 2010). The optical/nIR spectrum of the PWN is also modelled with a PL ($\alpha^{\rm PWN}_{\rm O,nIR}= 0.56 \pm 0.03$) which undershoots the extrapolation of the X-ray spectrum. This trend, so far observed only for the PSR\, J1124$-$5916 PWN, implies  a single break in the optical/IR--to--X-ray spectrum of the \psr\ PWN, at variance with the pulsar for which two breaks are required.  The optical/nIR spectrum of the PWN is also slightly flatter than the pulsar's, suggesting a possible spectral hardening at larger distance from the latter.
Like in the optical,  we could not find any obvious evidence of nIR spectral variations in the PWN on angular scales smaller than $\sim 1\arcsec$.  Only future observations with the \jwst\ will make it possible to extend the study of \psr\ and its PWN towards the mIR and search for a break in the pulsar spectrum at wavelengths up to $5 \mu m$, possibly establishing a similarity with the Crab and Vela pulsars. At the other end of the optical spectrum, multi-band observations in the nUV with the \hst, never performed so far, are required to search for the expected turnover between the optical and the X-rays.

\begin{acknowledgements}
RPM thanks  F. Primas (ESO) for her support during the \vlt\ Phase II proposal submission and D. Dobrzycka (ESO) for information on the \naco\ pipeline. We thank the anonymous referee for his/her constructive comments to our manuscript. A. S\l{}owikowska would like to acknowledges support from the Foundation for Polish Science grant FNP HOM/2009/11B, as well as from the Marie Curie European Reintegration Grant within the 7th European Community Framework
Programme (PERG05-GA-2009-249168). 
\end{acknowledgements}

\end{document}